# Securing Big Data from Eavesdropping Attacks in SCADA/ICS Network Data Streams through Impulsive Statistical Fingerprinting

Junaid Chaudhry[1], Uvais Qidwai[2] and Mahdi H. Miraz[3,4]

[1] Duja Inc., Perth, Australia
chaudhry@ieee.org
[2] Qatar University, Doha, Qatar
uqidwai@gmail.com
[3] The Chinese University of Hong Kong, Hong Kong
[4] Wrexham Glyndŵr University, UK
m.miraz@ieee.org

**Abstract.** While data from Supervisory Control And Data Acquisition (SCADA) systems is sent upstream, it is both the length of pulses as well as their frequency present an excellent opportunity to incorporate statistical fingerprinting. This is so, because datagrams in SCADA traffic follow a poison distribution. Although wrapping the SCADA traffic in a protective IPsec stream is an obvious choice, thin clients and unreliable communication channels make is less than ideal to use cryptographic solutions for security SCADA traffic. In this paper, we propose a smart alternative of data obfuscation in the form of Impulsive Statistical Fingerprinting (ISF). We provide important insights into our research in healthcare SCADA data security and the use of ISF. We substantiate the conversion of sensor data through the ISF into HL7 format and define policies of a seamless switch to a non HL7-based non-secure HIS to a secure HIS.

## 1    Introduction

Healthcare data has superior black market cyber value compared to financial data as noted by Trend Micro [1]. With an increasing number of personal healthcare records being digitised, for obvious reasons, the number of cyber-attacks on Health Information System has increased dramatically in the past few years [2].

The ISO/IEEE 11073 Personal Health Data (PHD) is a family of standards that enables medical/healthcare data to be exchanged between medical/healthcare/wellbeing devices and external computing systems. The Health Level Seven (HL7) has been providing data modelling and standardization support for almost 30 years. The successor of HL7 is the Fast Healthcare Interoperability Resources (FHIR) draft specification [3]. However, The FHIR specifications and its implementation are not within the scope of this paper. Rather, we chose to use HL7 due to its maturity. In fact, HL7 stays as the most widely accepted and adopted healthcare information sharing standard.

The ISO/IEEE 11073 and HL7 have successfully played a key role in enabling the medical devices to exchange data amongst themselves. Due to privacy and security issues, this data needs to be protected. Since medical devices are not typically computationally very proficient and the standards advocate lightweight communications, a cryptographic data security scheme is not an ideal solution to go ahead with [4]. In order to address this problem, we propose to use the statistical fingerprinting as data obfuscation method.

The Impulsive Statistical Fingerprinting (ISF) was used by Crotti et al. to classify the network traffic for anomaly detection [5]. When network traffic shows an atypical pattern and skewness from the "normal" traffic pattern, this is raised as an anomaly to the system [6]. We use the ISF to prevent leaf sensors from broadcasting the actual readings. Since the ISO/IEEE 11073 allows the flexibility to transmit metadata in the HL7 packet, we calculate the statistical model proposed in this paper to calculate the obfuscated values from sensors before translating the communication in the HL7 format.

The statistical model proposed in this paper assumes that the devices are already connected and configured in a network. An initial handshaking between PHDs and IP network provide the baseline values for mean calculation that will help in recalculating back to original values of the sensor data. From the results reported in [7] we learn that the PHDs send periodic metric report containing time stamp, and status value which reports device errors and alarms. The time stamps received from the devices helps keep both the TCP session [8] and time series data consistent. Using the statistical method proposed in [9], which assumes that the time series data is stationary, we consult to the history of the PHD values. We use the algorithm proposed in [9] and apply autoregressive integrated moving average models to find the best fit of the time series model to past values of the time series. An adversary, who is using a Man-in-The-Middle (MiTM) attack, needs to know considerable amount of time series data to predict the future values.

Following are the advantages of using statistical fingerprinting:
1) There is virtually no extra overhead in the existing infrastructure,
2) Adoption to the new solution is typically fairly straightforward and
3) The obfuscation of the patient data complies with privacy laws of both Australia [10] and Qatar [11], where the test-beds are planned for implementation and trial purposes.

## 2    Paper Organisation

Section III provides relevant definitions and background information on PHDs as well as the security and privacy goals. Section IV delivers the circumstantial account on the proposed method. Section V of the paper comprises a meticulous breakdown of the state-of-the-art in PHD security and privacy research. The concluding remarks have been discussed in section VI whereas the future works have been ventilated in section VII.

# 3 Background and Definitions

Recent advances in the embedded technologies and their usage in the field of medicine [12] have increased the benefits of integrated digital patient data to the highest levels. Seamless availability of medical data through interconnected systems has been made possible through the presence of standards [13]. These standards have evolved over time and have served as the key driving force in advent of PHD clusters at one/different HISs.

## 3.1 PHD Clusters and Inter/Intra Cluster Communication, Translation, and Addressing

According to ISO/IEEE 11073 standard [13], the Personal Healthcare Devices (PHDs) communicate through IEEE 11073 agents with the 11073 manager through a plethora of communication interfaces. These communication interfaces may range from IPv6 enabled Low powered Personal Area Networks (6LowPAN), Zigbee, Bluetooth, or 802.11x etc [14]. The point to point communication among the PHDs has been left on user settings i.e. as per needed. The devices communicate to the outside world through a hierarchy that includes device agents that are associated intern with the agent managers. This helps in scaling the device clusters in a better way.

## 3.2 Healthcare Data Security: Needs and Wants

There has been a tremendous growth in breaches into medical and healthcare data [1]. Healthcare data consists of patients' personal physical condition, financial information, private location information, images, videos, peers information, etc. Due to unavailability of legislation on collection of personal data in a HIS settings in the past and increasing availability of healthcare archives, there is a great amount of unfiltered data in healthcare databases. The IEEE 11073 assumes very little responsibility of privacy and security of user data. The standard assumes that each user agent communicates with only one manager and each of them is free to keep their own copy of the data. The presence of these copies makes the layered architecture proposed in IEEE 11073, a vulnerable data network architecture.

## 3.3 Data Obfuscation in Healthcare Information System

Since data is kept archived in its original form at different layers of communication, the risk of data loss and corruption increases exponentially. There is a growing need for a data security protocol which secures the data in transit and in archives. Because the IEEE 11073 encourages interoperability of heterogeneous devices and simplicity of the infrastructure, it is impractical to get the whole industry to agree on a set of encryption scheme. The PHDs have high diversity in computational power among themselves. Different encryption algorithms have different computational power [4]. We identify that there is a need for a lightweight data obfuscation scheme.

### 3.4    Statistical Fingerprinting

Due to the constant data stream that is periodically transmitted by the PHDs and inability of agreement to one encryption scheme, we propose to apply linear time series data aggregation algorithms to predict the next value [15] to calculate the autoregressive moving averages of the values that are generated by the PHDs. Using this statistical method, there is no need to change standards or manufacturer's device specifications.

## 4    The Proposed System

According to the IEEE 11073, the disease management devices fall in the 10400-10439 domain of standards. The PHD agents typically have limited capabilities i.e. RAM, ROM, CPU, etc., connection to a single PHD manager, limited resource in power, they are fixed in their configurations, and unreliable connection to the managers. On the other hand, PHD managers are higher in abstraction layer to the PHD agents with multiple connections, richer power resources, and higher processing capability. The figure 1 shows a typical PHD cluster scenario:

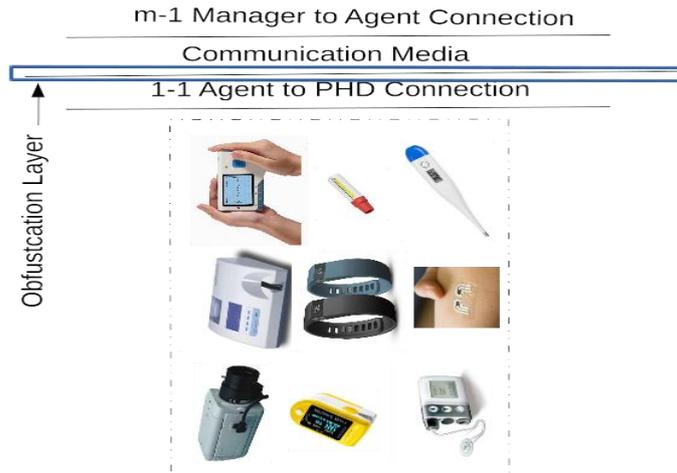

**Fig. 1.** The Organisation of PHDs, Agents, and Managers in an IEEE 11073 compliant cluster.

As shown in figure 1, the obfuscation layer that we propose in this paper resides between the agent and manager. We believe that the placement of obfuscation layer is critical considering the constraints posed by the device and the practice of data archiving as discussed in the previous section.

Let k be the PHD value at a given time $t_k$.

Where, we have an array of values that a PHD generates and those values are interrelated through a common source or by a common noise $k_i \in k[I]$. So the range of values that are generated from the PHD initially is the sample base for the statistical forecasting model that works on time series data. This series is generated by:

$$R_n = \sum_{i=0}^{i=n} K_i/t$$

.

Since it is well known, the variance, $\sigma^2 = (\sum_{i=0}^{i=o} \kappa_i - \mu)^2/n$. We can derive that if $K[I]$ be a numerically-valued discrete random variable with sample space $\Omega$ and distribution function $m(x)$.

Now, if $\sigma^2$ is added with the mean of the time series values, the original value can be obtained: $O_v = \sigma_2 + O[I]_{mean}$. The expected value $E(X)$ at PHD Agent is defined by

$$E(K[I]) = \sum_{x \in \Omega} x m(x)$$

. The proposed scheme shall be considered valid if and only if $t_k = (\gamma . E(X))/\beta$ .

Where $\gamma$ and $\beta$ are coefficients of consistency within the tolerance ranges of error.

### 4.1 The Testbed

We use the testbed presented in [16]: a testbed for monitoring the long term bedridden patients. The testbed consists of pressure sensors that generate events (pressure readings) which generate time series data & triggers the IEEE 11073 compliant video camera and can communicate to the PHD agent. There are 25 force sensitive sensors that are placed under the bedsheet in 5x5 strips manner. Each row of sensors is supported by the enamelled wiring and sown in a row into fabric for patient comfort. The sensors are arranged in the form of a matrix on the bed. The position of the sensors is adjustable according to the morphology and physiological features of the patient. The intensity of pressure/force on the sensors by a patient's posture on laying position in bed is measured by the voltage amplitude. The implementation of the PHD agent is performed on the Arduino board that is attached with wires to the sensors.

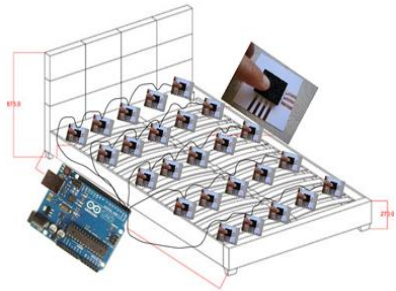

**Fig. 2.** Matrix of sensors in the long term bedridden patients and an Arduino board with PHD agents implementation and a host to the obfuscation layer algorithm.

The intensity of pressure/force on the sensors by patients' posture on laying position in bed is measured by voltage amplitude. The implementation of the PHD agent is performed on the Arduino board that is attached with wires to the sensors. The matrix of force sensors provide uniform data series. In order to achieve closeness to the real time environment, we introduce an event triggered digital camera. The data feed from the digital camera is routed to the IP network through the Arduino board.

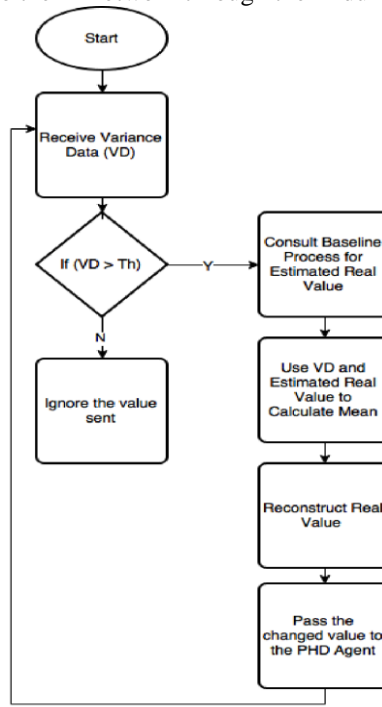

**Fig. 3.** The Flow chart of flow from the PHD to the agent and back.

In figure 4, we present the sequence flow diagram of the protocol proposed. We propose that the PHD transmits the baseline values at initial handshake. This is achieved after initial DHCP. After the acquisition of the baseline values, the PHD starts sending variance of the values with respect to the previous value transmitted. This way, the burst frequency is reduced which will result into lesser network traffic and if the sensor values stay the same, the network will carry packets with zero data values in it.

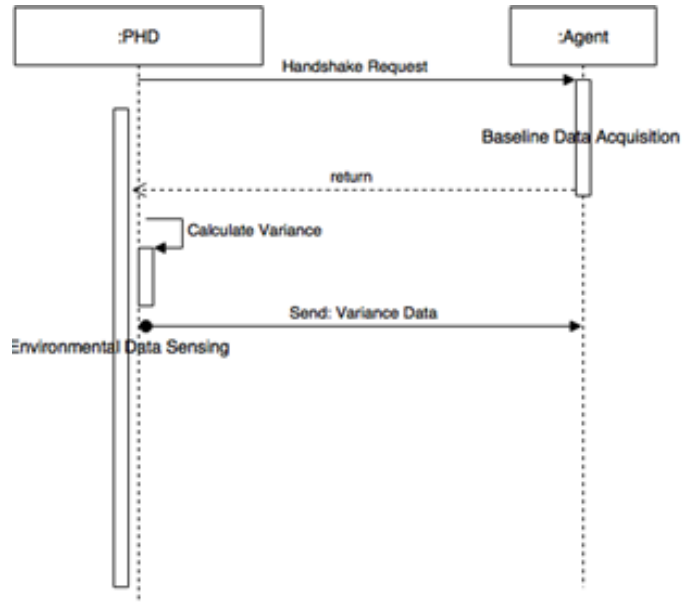

**Fig. 4.** The sequence diagram of the protocol proposed.

### 4.2 Multisensor Data Obfuscation through Statistical Fingerprinting

The continuous time series data that PHDs send through PHD agents to the PHD managers is in plain format in the conventional HIS. We propose that instead of sending sensor values, the variance from the previous value is sent. In [17], the algorithm for calculation of variance from an incoming data stream is reported. Using the stream of data carrying variance of sensor readings and mean from the baseline sensor readings, one can calculate the original sensor reading. This way, the data in transit stays obscure to the intruders.

The initial handshake in the system is based on trust. That is the devices that are newly configured to the system, share their 'native' sensor readings with the agents: residing on the Arduino. It is only after a sizable sample, in our case 50 samples, or native values are stored as demonstrated by the authors in [9]. An interesting debate on the usability of Kalmans filter and statistical methods are found in [9] which assisted us in choosing the statistical methods for this particular scenario.

The following notations describe the method of calculating the forecasted value for the time series under consideration using the statistical methods. This calculated value shall be used as the confidence value against packet injection attacks in a PHD networked environment. Although the statistical methods have their shortcomings [18], we believe that an alternative incorporation of input verification method shall greatly benefit the security in the PHD based networking environment.

### 4.3 Conversion of Obfuscated data into HL7 format

From [19], we learnt that conversion of data streams into HL7 format requires the policy engine that generates XML documents when provided with data source. In the current scenario, we have two series of data streams for the policy engine: 1- time series data to be kept as a baseline sample for the statistical model, 2- Variance data as time series data. We plan to implement a new version of the conversion software into the Weka tool [20].

### 4.4 Threat Model for the Proposed Scheme

In this section we discuss the threat model for the proposed scheme. As discussed earlier, that the ISO/IEEE 11037 standards encourages the transit of healthcare data through various networked devices. Mostly, this data is in plain text format. This data in transit, while unencrypted, is a serious security and privacy threat.

Let's assume that Alice is an attacker who wants to steal the healthcare data. In conventional environment, Alice will spoof network packets, or spoof IP of the PHD-Agent and request the latest sensor reading from the PHD. If the method proposed in this paper is adopted, the attacker will have the following scenarios to attack and steal the data.

***Through Packet Capture:*** If an attacker captures the packet that contains the PHD data, the attacker shall get variance from the data packet, which will not be of any use unless the attacker knows the mean of over the range of the values.

***Through Spoofing PHD-Agents IP address:*** This type of attack is typically classified as Man in The Middle (MiTM) attack. If Alice sends a decoy packet to the PHD agent and an association request, unless Alice has the statistical model, she cannot generate the mean from the baseline values and hence the attack will only be valid for the amount of time it takes to generate baseline values.

## 5 Empirical Analysis and Discussion

In our experiments, we extracted a sample set of 1400 readings from embedded temperature sensor hosted on the Adruino. The following diagram shows the dataset:

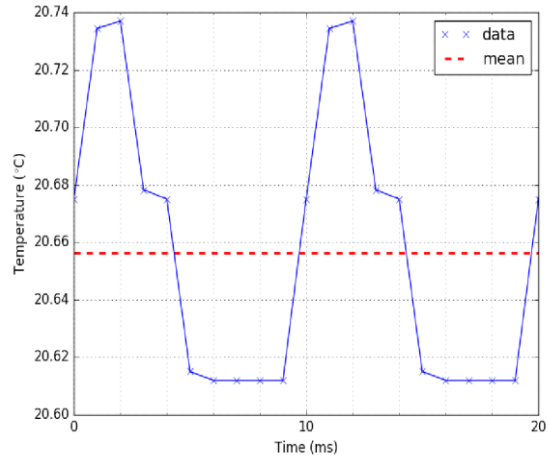

**Fig. 5.** The sample data set.

*Lemma 1: Whether we can constitute the original value from variance and mean?*

We extracted the mean and variance of the sample values which were 20.65627 and 0.00233 respectively. We determined that the reconstructed value was an exact match with the real value at the sensor. Whether this condition will still hold true for change in variance? - we rest this argument for the future work.

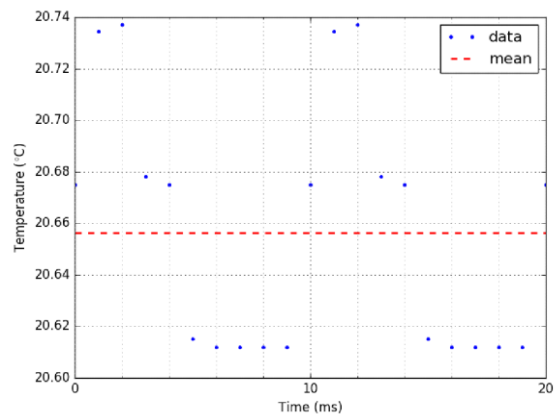

**Fig. 6.** Comparison of the Original Value at Sensor and Recalculation of Original Value at PHD Agent.

*Lemma 2: Whether the predicted value from the statistical model matches with the real value?*

In our observation, we noticed that the variance values for each sensor value predicted filled in perfectly in the gaps between the estimated value and the real value at the sensor. In future, we aim at investigating the effectiveness of the statistical finger-printing methods for different types of traffics. We also aim at testing these methods in different environments and investigate the effects of multi-sensor data fusion environments.

In past, we proposed the method in [19] that addresses conversion of the healthcare data into HL7 format. Using statistical fingerprinting, the volume of traffic is reduced by 66.75%. This is so because the statistical model requires two previous neighbouring values to serve as a reference point. The HL7 packet is going to hold additional statistical value within the packet to reflect the volume of the sample and variance of the values. This also reflects in the power consumption of the transmitting nodes which is not within the scope of this paper.

## 6    Related Works

Martinez *et al.* [21] present the implementation of an end-to-end standard-based patient monitoring solution. They demonstrate that their implementation is incompliant with X73 and EN13606 standards. However, data security in the system proposed by Martinez *et al.* rely on the security strengths of its components i.e. database security. The data in transit and cache stored at each level, according to the IEEE standard, stays vulnerable. The authors of [22] have enumerated the threat space for the medical devices. They go on to state that despite the vulnerabilities in the communication technologies, the data in transit is still vulnerable to attacks. Lee *et al.* [14] enumerate the interoperability challenges in the personal healthcare devices and also list that the data in transit as an issue that is unaddressed. In [23], Marc *et al.* present the case study of hospital image archive security on super clouds. They propose to use network services as the main computing power behind transfer and encryption/decryption of secure data. Delegation of AAA (Authentication, Authorization and Accounting) functions to the hypervisor might be considered rational but with increase in network volume, the idea of clouds of clouds might get in jeopardy. Zheng *et al.* [24] illustrate the application of blockchain technique in the PSN-based healthcare. The application of the blockchain in healthcare environment is unparalleled but since healthcare networks are hybrid in nature i.e. they consist of both thick and thin clients, blockchain might be overkill. An application of secure Internet of Things (IoT) in healthcare is proposed in [25]. On an IP-based network, on thin clients, the authors have proposed hashing messages and propose to use the hashing time as verification measure between the two processes. Since IoT is a de-centric concept, considerable amount of energies are invested in clock synchronisation. Timing-based process verification in absence of a/or two master clock(s) is risky. In [26], the authors have implemented a standard complaint prototype. They have discussed the plug and play capabilities derived from X73 standards. In compliance with ISO11037, the authors have not discussed the issues like security of data in transit, security of data in storage, device security, connection security, etc. They have relied on the inherent security features of TCP/IP.

Zhang *et al.* [27, 28] used statistical fingerprinting to detect peer-to-peer botnets through host-to-host traffic analysis. In [29], Trevor *et al.* is the RF fingerprinting for feature selection in Zigbee emissions and in [30], Sucki *et al.* use spatial fingerprinting to improve security in wireless networks. The statistical fingerprinting provides an in-vivo baseline to match the anomaly against, which is one of the most research areas in anomaly detection research. The foremost advantage of this non-crypto-based data obfuscation scheme proposed in this research paper is the lightweightness, portability, scalability, with very nominal overhead costs. Lee *et al.* [31] discussed several cases where anomaly detection is hard in health care networks because attackers mask their traffic with the normal traffic. In this situation, signature-based or traffic pattern based anomaly detection is near impossible. They emphasize on the use of verified transactions in the network which is an unrealistic assumption in large networks such that of health care networks. Zhenyu *et al.* [32] and in [33,34] addressed large-scale internet of things and its application in health care through fog computing. Having considered the rapid volumetric increase in the IoT technology, unless the underlying flaws of the supportive technologies are fixed, the cyber security is going to stay as a major cause of concern[35,36].

The ISO/IEEE 11073 Personal Health Data (PHD) is a family of standards that enables medical/healthcare data to be exchanged between medical/healthcare/wellbeing devices and external computing systems. The Health Level Seven (HL7) has been providing data modeling and standardization support for almost 30 years. The successor of HL7 is the Fast Healthcare Interoperability Resources (FHIR) draft specification [3]. However, The FHIR specifications and its implementation are not within the scope of this paper. Rather, we chose to use HL7 due to its maturity. In fact, HL7 stays as the most widely accepted and adopted healthcare information sharing standard.

## 7 Concluding Remarks

In this paper we presented our research using statistical fingerprinting before transformation of the healthcare data into HL7 format. We demonstrated that the statistical fingerprinting not only adds a semantic layer of security to the data, it can be performed locally at the data acquisition site without considerable overhead. The process of ISF is performed at the presentation stage of the data which means that the integrity of data is ensured. After the ISF is performed, the data is converted into HL7 format using the converter built in Java. If the data packet containing fingerprinted data is intercepted, an adversary will not be able to predict the value of the sensor unless s/he is physically present at the sensor site or has been present throughout the lifetime of the sensor sending the fingerprinted data. We believe that the research presented in this paper is of significant value because it advocates for seamless of HIS into a more secure HIS without considerable overhead or changes to the existing infrastructure.

## 8 Future Works

We aim at presenting a comprehensive comparative analysis of various encryption techniques, data obfuscation techniques, and the scheme proposed in this paper in an HIS setting. This tasks requires analysis of "key-driving variables" in a Health Information

System (HIS), foresighted strengths and weaknesses of cryptographic schemes, and development of context in order to choose the right data security scheme. We also plan to develop an adversarial model for the scheme proposed.